\documentstyle[preprint, aps]{revtex}

\begin{document}

\begin{center}
{\Large {\ {\bf Electron Spin Dynamics in Semiconductors without Inversion
Symmetry \\[0pt]
}} }{\large {Vadim I. Puller$^{1}$, Lev G. Mourokh$^{1}$, Anatoly Yu. Smirnov%
$^{2}$ and Norman J.M. Horing$^{1}$\\[0pt]
} }$^{1}${\it Department of Physics and Engineering Physics, Stevens
Institute of Technology, Hoboken, NJ 07030, USA \\[0pt]
}$^{2}${\it D-Wave Systems Inc., 320-1985 W.Broadway, Vancouver, B.C.,
Canada V6J 4Y3\\[0pt]
} \vspace*{0.2cm} {\bf Abstract}
\end{center}

{\footnotesize We present a microscopic analysis of electron spin dynamics
in the presence of an external magnetic field for non-centrosymmetric
semiconductors in which the D'yakonov-Perel' spin-orbit interaction is the
dominant spin relaxation mechanism. We implement a fully microscopic
two-step calculation, in which the relaxation of orbital motion due to
electron-bath coupling is the first step and spin relaxation due to
spin-orbit coupling is the second step. On this basis, we derive a set of
Bloch equations for spin with the relaxation times $T_{1}$ and $T_{2}$
obtained microscopically. We show that in bulk semiconductors without
magnetic field, $T_{1}=T_{2}$, whereas for a quantum well with a magnetic
field applied along the growth direction $T_{1}=T_{2}/2$ for any magnetic
field strength. \vspace*{0.5cm}}

Recent consideration of spin based quantum computation, optical switches,
magnetic memory cells, etc. [1] mandates an improved understanding of spin
dynamics and, in particular, the spin relaxation rate [2]. The most
promising materials for device purposes, III-V and II-VI compounds, have
been shown [3] to have spin relaxation rates dominated by the
D'yakonov-Perel' (DP) mechanism at moderate temperatures and low hole
concentrations, and the spin relaxation time is given by the following
semiphenomenological expression [4]: 
\begin{equation}
\frac{1}{\tau _{s}}=q\frac{\alpha ^{2}}{\hbar ^{2}\varepsilon _{g}}\tau
_{p}T^{3}  \label{DP}
\end{equation}
where $\alpha $ describes conduction band spin splitting due to lack of
inversion symmetry (for example $\alpha =0.07$ for GaAs), $\varepsilon _{g}$
is the band gap, $\tau _{p}$ is the average momentum relaxation time, and $T$
is the Kelvin temperature ($k_{B}=1$). The numerical coefficient $q$ depends
on the orbital scattering mechanism.

We develop a fully microscopic theory of spin relaxation in semiconductors,
based on a two step analysis of the relaxation process, corresponding to the
relaxation time hierarchy involved in (a) electron thermalization due to
dissipative bath action, and (b) spin relaxation. In the first stage of
solution, we can determine the relaxation rates and fluctuation
characteristics of electron orbital motion due to coupling to the bath. Spin
relaxation dynamics can be neglected in this stage.

Our analysis of the second stage proceeds with the spin relaxation process
due to spin-orbit interaction, wherein the orbital degrees of freedom are
considered as an effective heat bath, having its characteristics determined
in the first stage. A set of Bloch equations having two microscopically
determined relaxation times (a longitudinal relaxation time, $T_{1}$,
responsible for spin magnetic moment relaxation, and a transverse relaxation
time, $T_{2}$, responsible for decoherence) is derived in this second stage.
The method of Ref.[5] is employed in both stages of our analysis .

The orbital electron dynamics are determined from operator equations having
the form 
\begin{eqnarray}
\left( \frac{d}{dt}+\gamma _{0}\right) V_{x}(t)+\left( \omega _{c}+\delta
\right) V_{y}(t) &=&\xi _{x}(t),  \label{Velocity} \\
\left( \frac{d}{dt}+\gamma _{0}\right) V_{y}(t)-\left( \omega _{c}-\delta
\right) V_{x}(t) &=&\xi _{y}(t),  \nonumber
\end{eqnarray}
and 
\[
\left( \frac{d}{dt}+\gamma _{z}\right) V_{z}(t)=\xi _{z}(t), 
\]
where $V_{x}(t),V_{y}(t),V_{z}(t)$ are electron velocity operator components
($V_{x}=\left( p_{x}-m\omega _{c}y/2\right) /m$; $V_{y}=\left( p_{y}+m\omega
_{c}x/2\right) /m$; $V_{z}=p_{z}/m$; $\left[ V_{x},V_{y}\right] _{-}=-i\hbar
\omega _{c}/m$), and $\omega _{c}=\left| e\right| B/mc$ is the cyclotron
frequency. The electron-bath interaction determines the relaxation rates, $%
\gamma _{0}$,$\gamma _{z}$, the frequency shift, $\delta $, and the
fluctuation sources, $\xi _{x}(t),\xi _{y}(t),\xi _{z}(t)$, involved in the
Eq. (\ref{Velocity}). The Fourier transforms of the velocity correlation
functions are given by 
\begin{equation}
\left\langle \frac{1}{2}\left[ V_{x}(\omega );V_{x}\right] _{+}\right\rangle
=\left\langle \frac{1}{2}\left[ V_{y}(\omega );V_{y}\right]
_{+}\right\rangle =\frac{K_{\bot }(\omega )}{2}\left( \frac{1}{\left( \omega
-\omega _{c}\right) ^{2}+\gamma _{0}^{2}}+\frac{1}{\left( \omega +\omega
_{c}\right) ^{2}+\gamma _{0}^{2}}\right) ,  \label{xx}
\end{equation}
\begin{equation}
\left\langle \frac{1}{2}\left[ V_{x}(\omega );V_{y}\right] _{+}\right\rangle
=-\left\langle \frac{1}{2}\left[ V_{y}(\omega );V_{x}\right]
_{+}\right\rangle =\frac{K_{\bot }(\omega )}{2i}\left( \frac{1}{\left(
\omega -\omega _{c}\right) ^{2}+\gamma _{0}^{2}}-\frac{1}{\left( \omega
+\omega _{c}\right) ^{2}+\gamma _{0}^{2}}\right) ,  \label{xy}
\end{equation}
and 
\begin{equation}
\left\langle \frac{1}{2}\left[ V_{z}(\omega );V_{z}\right] _{+}\right\rangle
=\frac{K_{z}(\omega )}{\omega ^{2}+\gamma _{z}^{2}},  \label{zz}
\end{equation}
where 
\begin{equation}
K_{\bot }(\omega )=\int d\left( t-t_{1}\right) e^{i\omega \left(
t-t_{1}\right) }\left\langle \frac{1}{2}\left[ \xi _{x}(t),\xi _{x}(t_{1})%
\right] _{+}\right\rangle =\int d\left( t-t_{1}\right) e^{i\omega \left(
t-t_{1}\right) }\left\langle \frac{1}{2}\left[ \xi _{y}(t),\xi _{y}(t_{1})%
\right] _{+}\right\rangle ,
\end{equation}
\[
K_{z}(\omega )=\int d\left( t-t_{1}\right) e^{i\omega \left( t-t_{1}\right)
}\left\langle \frac{1}{2}\left[ \xi _{z}(t),\xi _{z}(t_{1})\right]
_{+}\right\rangle , 
\]
and $\left[ ...,...\right] _{+}$ denotes the anticommutator.

In the second stage we analyze spin relaxation due to the DP interaction
between spin and electron orbital motion. The corresponding interaction
Hamiltonian is given by [4] 
\begin{equation}
H_{DP}=-\sigma _{x}Q_{x}(t)-\sigma _{y}Q_{y}(t)-\sigma _{z}Q_{z}(t),
\end{equation}
where 
\begin{eqnarray}
Q_{x}(t) &=&-\lambda V_{x}(t)\left( V_{y}^{2}(t)-V_{z}^{2}(t)\right) ,\text{ 
}Q_{y}(t)=-\lambda V_{y}(t)\left( V_{z}^{2}(t)-V_{x}^{2}(t)\right) ,
\label{Q} \\
Q_{z}(t) &=&-\lambda V_{z}(t)\left( V_{x}^{2}(t)-V_{y}^{2}(t)\right) , 
\nonumber
\end{eqnarray}
and 
\begin{equation}
\lambda =\frac{\alpha m^{3/2}}{2\sqrt{2\varepsilon _{g}}}.
\end{equation}
Employing \ a second application of the method of Ref. [5] we consider the
orbital dynamics in the role of an effective heat bath and obtain a set of
Bloch equations for the average spin projections as 
\begin{eqnarray}
\frac{d}{dt}\left\langle \sigma _{x}(t)\right\rangle &=&-\frac{\left\langle
\sigma _{x}(t)\right\rangle }{T_{2}}-\left( \omega _{B}+\delta _{x}\right)
\left\langle \sigma _{y}(t)\right\rangle ,  \label{Bloch} \\
\frac{d}{dt}\left\langle \sigma _{y}(t)\right\rangle &=&\left( \omega
_{B}+\delta _{y}\right) \left\langle \sigma _{x}(t)\right\rangle -\frac{%
\left\langle \sigma _{y}(t)\right\rangle }{T_{2}},  \nonumber \\
\frac{d}{dt}\left\langle \sigma _{z}(t)\right\rangle &=&\frac{\sigma
_{z}^{0}-\left\langle \sigma _{z}(t)\right\rangle }{T_{1}},  \nonumber
\end{eqnarray}
where $\sigma _{x},\sigma _{y},\sigma _{z}$ are the Pauli matrices and $%
\sigma _{z}^{0}=-\tanh \left( \hbar \omega _{B}/2T\right) $ is the
equilibrium $z$-component of spin, $\omega _{B}=g\mu _{B}B/\hbar $, $\mu
_{B}=\left| e\right| \hbar /2m_{0}c$ is the Bohr magneton, and the $g$%
-factor depends on material (it is $-0.44$ for GaAs). Our microscopic
determination of the relaxation times $T_{1}$ and $T_{2}$ yields 
\begin{equation}
\frac{1}{T_{1}}=\frac{4}{\hbar ^{2}}\left( S_{xx}(\omega
_{B})+iS_{xy}(\omega _{B})\right) ,  \label{T1}
\end{equation}
and 
\begin{equation}
\frac{1}{T_{2}}=\frac{2}{\hbar ^{2}}\left( S_{xx}(\omega
_{B})+iS_{xy}(\omega _{B})+S_{zz}(0)\right) ,  \label{T2}
\end{equation}
where $S_{xx}(\omega _{B}),S_{xy}(\omega _{B})$, and $S_{zz}(0)$ are the
Fourier transforms of \ the correlation functions of the variables $%
Q_{j}(t), $ ($j=x,y,z$) of Eq. (\ref{Q}): 
\begin{equation}
S_{ij}(\omega )=\int d\left( t-t_{1}\right) e^{i\omega \left( t-t_{1}\right)
}\left\langle \frac{1}{2}\left[ Q_{i}(t),Q_{j}(t_{1})\right]
_{+}\right\rangle .
\end{equation}
Eq. (\ref{Q}) implies that the spectral functions $S_{ij}(\omega )$ are
averages of the sixth power of the electron velocity component operators,
and employing Wick's theorem, we obtain 
\begin{eqnarray}
S_{xx}(\omega ) &=&S_{yy}(\omega )=\lambda ^{2}\left\langle \frac{1}{2}\left[
V_{x}(\omega );V_{x}\right] _{+}\right\rangle \left[ \left( \left\langle
V_{z}^{2}\right\rangle -\left\langle V_{x}^{2}\right\rangle \right)
^{2}+4\left\langle V_{x}V_{y}\right\rangle \left\langle
V_{x}V_{y}\right\rangle \right] + \\
&&+\lambda ^{2}\int \frac{d\omega _{1}}{2\pi }\int \frac{d\omega _{2}}{2\pi }%
\Xi \left( \omega _{1},\omega _{2},\omega -\omega _{1}-\omega _{2}\right)
\cdot  \nonumber \\
&&\cdot \left\{ 2\left\langle \frac{1}{2}\left[ V_{x}(\omega _{1});V_{x}%
\right] _{+}\right\rangle \left\langle \frac{1}{2}\left[ V_{x}(\omega
_{2});V_{x}\right] _{+}\right\rangle \left\langle \frac{1}{2}\left[
V_{x}(\omega -\omega _{1}-\omega _{2});V_{x}\right] _{+}\right\rangle
+\right.  \nonumber \\
&&2\left\langle \frac{1}{2}\left[ V_{x}(\omega _{1});V_{x}\right]
_{+}\right\rangle \left\langle \frac{1}{2}\left[ V_{z}(\omega _{2});V_{z}%
\right] _{+}\right\rangle \left\langle \frac{1}{2}\left[ V_{z}(\omega
-\omega _{1}-\omega _{2});V_{z}\right] _{+}\right\rangle -  \nonumber \\
&&\left. -4\left\langle \frac{1}{2}\left[ V_{x}(\omega _{1});V_{x}\right]
_{+}\right\rangle \left\langle \frac{1}{2}\left[ V_{x}(\omega _{2});V_{y}%
\right] _{+}\right\rangle \left\langle \frac{1}{2}\left[ V_{x}(\omega
-\omega _{1}-\omega _{2});V_{y}\right] _{+}\right\rangle \right\} , 
\nonumber
\end{eqnarray}
\begin{eqnarray}
S_{xy}(\omega ) &=&-S_{yx}(\omega )=\lambda ^{2}\left\langle \frac{1}{2}%
\left[ V_{y}(\omega );V_{x}\right] _{+}\right\rangle \left[ \left(
\left\langle V_{z}^{2}\right\rangle -\left\langle V_{x}^{2}\right\rangle
\right) ^{2}+4\left\langle V_{x}V_{y}\right\rangle \left\langle
V_{x}V_{y}\right\rangle \right] + \\
&&+\lambda ^{2}\int \frac{d\omega _{1}}{2\pi }\int \frac{d\omega _{2}}{2\pi }%
\Xi \left( \omega _{1},\omega _{2},\omega -\omega _{1}-\omega _{2}\right)
\cdot  \nonumber \\
&&\cdot \left\{ 2\left\langle \frac{1}{2}\left[ V_{y}(\omega _{1});V_{x}%
\right] _{+}\right\rangle \left\langle \frac{1}{2}\left[ V_{y}(\omega
_{2});V_{x}\right] _{+}\right\rangle \left\langle \frac{1}{2}\left[
V_{y}(\omega -\omega _{1}-\omega _{2});V_{x}\right] _{+}\right\rangle
+\right.  \nonumber \\
&&+2\left\langle \frac{1}{2}\left[ V_{y}(\omega _{1});V_{x}\right]
_{+}\right\rangle \left\langle \frac{1}{2}\left[ V_{z}(\omega _{2});V_{z}%
\right] _{+}\right\rangle \left\langle \frac{1}{2}\left[ V_{z}(\omega
-\omega _{1}-\omega _{2});V_{z}\right] _{+}\right\rangle -  \nonumber \\
&&\left. -4\left\langle \frac{1}{2}\left[ V_{y}(\omega _{1});V_{x}\right]
_{+}\right\rangle \left\langle \frac{1}{2}\left[ V_{x}(\omega _{2});V_{x}%
\right] _{+}\right\rangle \left\langle \frac{1}{2}\left[ V_{x}(\omega
-\omega _{1}-\omega _{2});V_{x}\right] _{+}\right\rangle \right\} , 
\nonumber
\end{eqnarray}
and 
\begin{eqnarray}
S_{zz}(\omega ) &=&\lambda ^{2}\int \frac{d\omega _{1}}{2\pi }\int \frac{%
d\omega _{2}}{2\pi }\Xi \left( \omega _{1},\omega _{2},\omega -\omega
_{1}-\omega _{2}\right) \cdot  \nonumber \\
&&\cdot 4\left\langle \frac{1}{2}\left[ V_{z}(\omega _{1});V_{z}\right]
_{+}\right\rangle \left\{ \left\langle \frac{1}{2}\left[ V_{x}(\omega
_{2});V_{x}\right] _{+}\right\rangle \left\langle \frac{1}{2}\left[
V_{x}(\omega -\omega _{1}-\omega _{2});V_{x}\right] _{+}\right\rangle
-\right.  \nonumber \\
&&\left. -\left\langle \frac{1}{2}\left[ V_{x}(\omega _{2});V_{y}\right]
_{+}\right\rangle \left\langle \frac{1}{2}\left[ V_{x}(\omega -\omega
_{1}-\omega _{2});V_{y}\right] _{+}\right\rangle \right\} ,  \nonumber
\end{eqnarray}
where 
\begin{eqnarray}
\Xi \left( \omega _{1},\omega _{2},\omega _{3}\right) &=&1+\tanh \left( 
\frac{\hbar \omega _{1}}{2T}\right) \tanh \left( \frac{\hbar \omega _{2}}{2T}%
\right) +  \nonumber \\
&&+\tanh \left( \frac{\hbar \omega _{1}}{2T}\right) \tanh \left( \frac{\hbar
\omega _{3}}{2T}\right) +\tanh \left( \frac{\hbar \omega _{2}}{2T}\right)
\tanh \left( \frac{\hbar \omega _{3}}{2T}\right)
\end{eqnarray}
and $\left\langle V_{j}^{2}\right\rangle =\left\langle \frac{1}{2}\left[
V_{j}(t);V_{j}(t)\right] _{+}\right\rangle =\int \frac{d\omega }{2\pi }%
\left\langle \frac{1}{2}\left[ V_{j}(\omega );V_{j}\right] _{+}\right\rangle
,$ $\left\langle V_{x}V_{y}\right\rangle =\left\langle \frac{1}{2}\left[
V_{x}(t);V_{y}(t)\right] _{+}\right\rangle =\int \frac{d\omega }{2\pi }%
\left\langle \frac{1}{2}\left[ V_{x}(\omega );V_{y}\right] _{+}\right\rangle 
$

Any model of a bath (phonons, random impurities, etc.) can be accommodated
in this general formulation. Even without specifying the nature of the bath,
some peculiarities of spin dynamics can be identified. In particular, in the
absence of a magnetic field we obtain $T_{1}=T_{2}$($=\tau _{s}$ of Eq.(\ref
{DP}) with the replacement $q\tau _{p}\rightarrow 1/\gamma _{z}$), as the
zero field limit. However, in the presence of a magnetic field $T_{1}\neq
T_{2}$. For the case of strong confinement along the z-axis (quantum well
with magnetic field in the growth direction) we have $S_{zz}(\omega )=0$ and 
$T_{1}=T_{2}/2$ regardless of \ the magnetic field strength.

In summary, we have derived Bloch equations for electron spins in
non-centrosymmetric semiconductors on a fully microscopic basis in the
presence of a magnetic field. Explicit expressions for the spin relaxation
times have been derived and analyzed using the D'yakonov-Perel' spin-orbit
interaction mechanism.

V.I.P., L.G.M. and N.J.M.H. gratefully acknowledge support from the
Department of Defense, DAAD 19-01-1-0592. V.I.P. also thanks the organizing
committee of MS+S2002 conference for financial support for his participation.

\begin{center}
{\bf References}
\end{center}

[1] S.A. Wolf, D.D. Awschalom, R.A. Buhrman, J.M. Daughton, S. von
Moln\'{a}r, M.L. Roukes, A.Y. Chtchelkanova, D.M. Treger, Science {\bf 294},
1488 (2001); G.A. Prinz, Science {\bf 282}, 1660 (1998); Y. Nishikawa, A.
Tackeuchi, S. Nakamura, S. Muto, and N. Yokoyama, Appl. Phys. Lett. {\bf 66}%
, 839 (1995).

[2] W.H. Lau, J.T. Olesberg, and M.E. Flatt\'{e}, Phys. Rev. B {\bf 64},
161301(R) (2001); A. Bournel, P. Dolfus, E. Cassan, and P. Hesto, Appl.
Phys. Lett. {\bf 77} 2346 (2000); D.M. Frenkel, Phys. Rev. B {\bf 43} 14228
(1991); P.H. Song and K.W. Kim, arXiv:cond-mat/0111076 (2001).

[3] J.M. Kikkawa and D.D. Awschalom, Phys. Rev. Lett. {\bf 80}, 4313(1998);
T.F. Bogges, J.T. Olesberg, C. Yu, M.E. Flatt\'{e}, and W.H. Lau, Appl.
Phys. Lett. {\bf 77}, 1333 (2000); B. Beschoten, E. Johnston-Halperin, D.K.
Young, M. Poggio, J.E. Grimaldi, S. Keller, S.P. DenBaars, U.K. Mishra, E.L.
Hu, and D.D. Awschalom, Phys. Rev. B {\bf 63} 121202(R) (2001).

[4] M.I. D'yakonov and V.I. Perel', Sov. Phys. JETP {\bf 33}, 1053 (1977); 
{\it Optical Orientation, Modern Problems in Condensed Matter Science},
edited by F. Meier and B.P. Zakharchenya (North-Holland, Amsterdam, 1984),
Vol. 8.

[5] G.F. Efremov and A.Yu. Smirnov, Sov. Phys. JETP {\bf 53}, 547 (1981);
L.G. Mourokh and S.N. Zheltov, Physica B {\bf 228}, 305 (1996); A.Yu.
Smirnov, Phys. Rev. E, {\bf 56}, 1484 (1997); G. Rose and A.Yu. Smirnov J.
Phys.: Condens. Matter {\bf 13}, 11027 (2001).

\end{document}